# Aging effects in the COMPASS hybrid GEM-Micromegas pixelized detectors

Damien Neyret[1,*], Philippe Abbon[1], Marc Anfreville[1,†], Vincent Andrieux[2], Yann Bedfer[1], Dominique Durand[1], Sébastien Herlant[1], Nicole d'Hose[1], Fabienne Kunne[1], Stephane Platchkov[1], Florian Thibaud[1], Michel Usseglio[1], Maxence Vandenbroucke[1]

[1] CEA IRFU, Université Paris-Saclay, 91191 Gif sur Yvette Cedex, France

[2] University of Illinois at Urbana-Champaign, Dept. of Physics, Urbana, IL 61801-3080, USA

## Abstract

Large-size hybrid and pixelized GEM-Micromegas gaseous detectors (40x40 cm² active area) were developed and installed in 2014 and 2015 for the COMPASS2 physics program which started at the same time. That program involved in particular two full years of Drell-Yan studies using a high-intensity pion beam on a thick polarized target. Although the detectors were placed behind a thick absorber, they were exposed to an important flux of low energy neutrons and photons. The detectors were designed to drastically reduce the discharge rate, a major issue for non-resistive Micromegas in high hadron flux, by a factor of more than 100 compared to the former ones. A hybrid solution was chosen where a pre-amplifying GEM foil is placed 2 mm above the micromesh electrode. A pixelized readout was also added in the center of the detector, where the beam is going through, in order to track particles scattered at very low angles. The combination of the hybrid structure and the pixelized central readout allowed the detector to be operated in an environment with particle flux above 10 MHz/cm² with very good detection efficiencies and spatial resolution. The performance has remained stable since 2015 in terms of gain and resolution, showing the interest of hybrid structures associating a GEM foil to a Micromegas board to protect gaseous detectors against discharges and aging effects

**Keywords**: Micromegas hybrid detectors, Micro-patterned Gaseous Detectors (MPGD), Detector efficiency, Spatial resolution, Detector aging study, COMPASS experiment

## 1. Introduction

The first detectors based on the Micromegas technology [1] were developed and optimized in the late 90's. They were used in the COMPASS experiment at CERN as trackers for particles scattered at small angle [2]. They were placed right after the target, where the particle flux is the largest, and designed to track charged particles at a radial distance from the beam from 2.5 up to 20 cm, on a surface of 40x40 cm². Twelve planes were installed on 3 stations, consisting of 4 detectors in 4 different orientations X, Y, U and V (respectively 0°, 90° and ±45°). They were using a gas mixture of Neon with 10% ethane and 10% of $CF_4$. These detectors were installed in 2001-2002 at the very beginning of the experiment, and showed excellent performance with muon beams in terms of efficiency and spatial and time resolutions. When used with hadron beams the relatively high sparking rate of the Micromegas detectors required to reduce the gain and to remove the $CF_4$ component from the gas mixture, leading to a slight decrease of the detector performance.

In anticipation for the second phase of the COMPASS experiment [3] [4], an upgrade of the Micromegas detectors was necessary with two main objectives: to increase the ability to operate in high hadron flux, and to make the new detectors active in their center, unlike in the first generation. An R&D was pursued in 2008-2013 in order to reach these objectives. New pixelized hybrid Micromegas detectors [5] were developed, built and installed in the COMPASS spectrometer in 2014-2015, with excellent performance. These detectors have been used on the COMPASS experiment until its termination at end of 2022. They are used since 2023 on the AMBER experiment [6].

---

\*   Corresponding author, email: damien.neyret@cea.fr

†   Deceased

## 2. Discharge reduction with hybrid structure

The impact of discharges is a recurrent issue for Micromegas detectors. The discharges appear when highly ionizing particles are impinging on the detector, creating in the amplification gap a too large charge density which reaches the Raether limit and thus inducing a conductive tube of ionized gas between the mesh and the anode plane. The consequence is a spark across the amplification gap which lasts until the voltage difference drops to almost zero. In order to avoid forcing the mesh potential to go to 0 V, in the COMPASS Micromegas the anode strips were decoupled by a 220 pF capacitance and an 1 MΩ resistor, letting the strips meet the mesh voltage with a much lower capacitance value, and thus a lower amount of charges involved in the discharge. Several ideas were developed to overcome the effect of the discharges, mostly based on a resistive layer placed above the anode plane [7] [8], which reduces the amplitude of the discharge by limiting the capacitance of the readout board area involved in the process. The effects of the discharge, like the variation of mesh voltage, is therefore very limited and barely visible.

A hybrid structure was studied in parallel, combining a regular Micromegas board with a preamplifying GEM foil [9] placed above the mesh and separated by a transfer gap of 1 or 2 mm (Figure 1). This concept was partially studied in the early 2000s [10] but never used in any application before the present project. By spreading the preamplified charges in the transfer gap, and reducing the gain of the Micromegas stage, the probability to reach the Raether limit and thus to trigger a discharge is expected to decrease by a few orders of magnitude [11].

Small hybrid prototypes of 6x10 cm² surface, as well as resistive prototypes, were tested in 2009 and 2010 in muon and hadron beams at CERN PS and SPS [12] [13] [14], in preparation for the COMPASS2 and CLAS12 at Jefferson lab projects [15]. Hybrid prototypes showed discharges with unchanged amplitudes, but with rates decreased by factors of 10 to 100 for a given gain, depending on the gain of the GEM foil (Figure 2), and hence fulfilling the requested performance. Resistive structures were also tested in the same beam campaigns, with different designs (resistive strips on insulator, resistive plane on insulator, resistive strips directly on electrodes, buried resistor structure with chicanes). Some of these prototypes gave promising results, and the resistive strips on insulator structure was finally chosen for the CLAS12 project. However some concerns appeared for the COMPASS2 application with resistive planes and strips as the particle flux may be concentrated in the center of the detectors, which may lead to variations of the resistive layer voltage, and thus of the gain, between the center and the edge of the detector. Buried resistor structure, which was not affected by this effect, was studied with a full scale 40x40 cm² prototype which gave promising results [16], but such a structure turned out to be too complicated for a mass production. Given the good performance of hybrid structure and the lack of valid resistive solution for our application, the hybrid structure was finally selected for this project.

## 3. New detectors with pixel readout in the detector center

In COMPASS, the muon beam can reach intensities up to $5.10^7$ μ/s and covers an area of a few cm². The flux in the center of the detectors can reach a level in the order of 100 kHz/mm². A readout with 400 μm-wide strips in the central area would lead to hit rates larger than 500 kHz/channel which would generate an inefficiency in the order of 10% due to electronics occupancy. To activate the central area of the new detectors it was necessary to decrease the hit rate by channel in order to keep the electronics occupancy below a few percent. The classical solution to pave the readout plane with square pixels was not optimal, due to the loss in spatial resolution it would have implied in the direction perpendicular to strips. In order to keep the number of channels reasonably low the square pixels would have a size of the order of 1 mm, to be compared with strip pitch of 400 μm. A better solution based on rectangular pixels with a width of 400 μm, identical to the strip pitch, and a length of a few mm was finally chosen (Figure 3). With a surface of 1 mm², the central pixels would reach an occupancy below 200 kHz for the highest beam intensities, assuming that a particle generates signals on slightly more than 2 pixels in average.

The final hybrid detectors were designed with this pixel readout on the center, on an almost circular area of a 5 cm diameter, and strips readout on the periphery, to cover the same 40x40 cm² surface as the original ones [16]. The gas mixture was the same as the former detectors. Electronic readout cards based on the APV ASIC [17] were produced to read Micromegas signals. The APVs were tuned to amplify and shape the signals with a peaking time of around 150 ns and an occupancy time around 400 ns. 3 samples are recorded per trigger in order to measure the amplitude and the time of the particle hits. Noise levels after common mode noise corrections are measured at values roughly around 600 e⁻. Overall performance of the detectors was measured with prototypes before the final installation [18] [16].

## 4. COMPASS conditions between 2014 and 2022

Two different types of runs took place between 2015 and 2018. The first type, called DVCS run [3] [19], used a medium-intensity muon beam (~1.4x10$^7$ µ/s) on a light liquid hydrogen target, while the second type, Drell-Yan run [20] [21], used a high-intensity pion beam (~7x10$^7$ p/s) on several consecutive thick targets: 110 cm polarized ammonia, 7 cm aluminium and 150 cm of tungsten. In 2022 a transversity run [22] [23] took place using a high-intensity muon beam (~5x10$^7$ µ/s) on a thick $^6$LiD polarized target. Runs with muon beams led to low particle flux on the detectors, except at the beam crossing spot where the local particle density could be as high as 100 kHz/mm².

During the Drell-Yan runs a thick concrete+alumina+steel absorber was placed around the tungsten target to stop the beam and the high energy hadrons emitted by the targets (Figure 4). However a large amount of low energy secondary neutrons and photons were going out of the absorber to the Micromegas detectors. A 2 cm thick lithium shield foil was added in 2018 after the absorber to stop part of the outgoing neutrons.

## 5. Radiation levels during the 2015 and 2018 Drell-Yan runs

Studies were done in 2011 and 2015 by Angelo Maggiora et al. [24] to estimate the impact of low-energy neutron and photon radiations emitted by the 2 m thick absorber to the polarized target, to the Micromegas detectors and from the point of view of radiation protection. These studies were based on FLUKA [25] simulations using version 2011.2c.5 to estimate the dose values on the different materials per primary particle. The radiation levels were estimated at the first Micromegas station with and without the additional lithium foil to reflect the 2015 and 2018 situations. Values of 2.1 photons and 1.7 neutrons per pion beam particle were estimated for 2015, and 2.2 photons and 1.38 neutrons per pion beam particle for 2018 with the additional lithium foil. The energies of the photons were roughly comprised between 100 keV and 10 MeV, and between 100 keV and 300 MeV for the neutrons (Figure 5). The particle were almost evenly distributed over the detector surface, with a ratio of 4 to 1 of the particle density between the center and the external parts. The total integrated irradiation on the Micromegas first station for 2015 and 2018 summed together was estimated to be 3.4x10$^{14}$ neutrons and 5x10$^{14}$ photons.

## 6. Micromegas electrode currents with high flux hadron beam

The large flux of secondary particles coming from the absorber during the Drell-Yan runs induced large currents on the Micromegas micromesh, due to the amplification processes. Historically, to estimate discharge rates one would study the mesh currents reported by the high-voltage power supply (CAEN A1821N) and count events with a sudden increase of a few hundreds of nA. However, this method was only working as the amplification current for the historical detector was much lower than the discharge one, at the level of a few tens of nA (Figure 6 left). During the Drell-Yan run and with the new hybrid detectors active in the center, the amplification current was much larger, up to 2 µA for the first station, the closest from the absorber (Figure 6 right). The expected discharge currents would be at the same level as the current fluctuations mainly due to beam intensity variations, and then barely visible. As no other direct discharge detection method was available, the discharge rate was not measured, and the impact of discharge could only be evaluated on the detector performance, in particular the detector efficiencies.

## 7. Detector performance

Detector pseudo-efficiencies are determined from the number of times a given detector reports a hit at the crossing position of a track reconstructed in the COMPASS spectrometer, corrected from the impact of pile-up contributions. The hits are searched within a route of a width of 3 times the detector resolution, so around 300 µm. This method is slightly biased as the studied detector is included in the track reconstruction algorithm, but in practice, with 12 Micromegas detector planes involved plus the other COMPASS detectors, the measured pseudo-efficiencies are close to the non-biased ones.

Efficiencies were measured for the various run conditions. Intrinsic detector efficiencies were specifically measured during the 2015 Drell-Yan run, but using low intensity muon beams which does not degrade the detector performance measurement: low discharge rate due to low ionization rate of muons, low multiplicity of tracks making them easy to reconstruct, and low detector channel occupancy. Values around 99% were measured for the first station in such conditions (Figure 7 left) for both the peripheral strip and the central pixel areas. On the other hand, efficiencies measured in high intensity pion beam for the same station 1 showed values reduced by a few percents (97-98%, Figure 7 center). Efficiencies were also extracted for the station 3 farther from the absorber, and with a lower low-energy particle flux due to absorption in the detectors placed upstream. In the strip area they were better by 1% with values of 98%, a non-significant increase. In the pixel area the values were approximately the same. Efficiencies were also measured during the DVCS runs (Figure 7 right), with medium intensity muon beam and without absorber, giving values between 98 and 99% for both strip and pixel area. However due to the light target and the low scattering angle observed for muons, tracks are mostly populating the pixel area and the strip area around the detectors center, so it is not representative of the efficiency of the whole strip area.

Spatial resolutions were determined by measuring the residual distance between the particle hit reconstructed from the detector data, and the impact point of the reconstructed track on the same detector. The resolutions are slightly biased the same way as efficiency measurements, as the studied detector was included in the track reconstruction. DVCS data with medium flux muon beam gave excellent residual values between 70 and 90 μm for the station 1 (Figure 8 left). These values are slightly over estimated since the much smaller intrinsic resolution of the tracks was not taken into account. For low intensity muon beam during Drell-Yan runs the performance is not as good, with residual values between 83 to 94 μm in both areas (Figure 8 center). During the Drell-Yan high intensity pion beam period resolutions were largely degraded with values between 110 and 130 μm (Figure 8 right). The degradation is attributed to the high flux of low-energy particles which add low amplitude random signals.

Detector time resolutions were also measured for the three beam conditions with similar values between 12 and 14 ns.

## 8. Performance in 2022

Detector performance were remeasured during the 2022 run with an high-intensity muon beam, to study possible aging effects after the integrated irradiation of 2014-2018. During this run the particles were mostly illuminating the pixel area and the strip area just around, covering a diameter of ~20 cm. Moreover the particle flux in the center of the pixel area reached large values at the level of 100 kHz/mm² over a surface of a few cm².

Efficiencies of 97-98% were measured for the strip areas. They were slightly smaller for the pixel area (95 to 96%), and even degraded at the beam position with values below 90% (Figure 9 left). The degradation is due to the large electronic occupancy of the pixel channels in this spot, leading to an overestimation of the correction for the pile-up contribution. Without this bad correction the center spot shows same efficiencies as the other parts of the detector, with values in the 97-98% range (Figure 9 center). Spatial resolution measurements show residual values between 80 to 90 μm, equivalent to the ones measured during the DVCS run (Figure 9 right). Time resolution are slightly larger than during the 2015-18 period with values between 15 to 20 ns, which can be explained by the large particle flux.

## 9. Conclusions

The pixelized hybrid Micromegas detectors were installed in 2014-15 in the COMPASS setup, and used under various run conditions between 2014 and 2022. They are still being used to date (2023-24) by the AMBER experiment. These detectors allowed to reduce by a factor larger than 100 the discharge rate, even though the detector centers are no longer blind as they were in the former Micromegas detectors.

The largest irradiation took place in 2014-2015 and 2018 during the Drell-Yan runs, where a high intensity pion beam was fully stopped in an absorber, which was emitting a large flux of low energy photons and neutrons. The other runs were based on muon beams on light targets which were delivering particle flux mainly on the detector centers. Detector performance were measured in the different beam conditions. The outcomes are that large channel occupancies have a temporary impact on the efficiency measurement, due to the pile-up contribution of non-correlated hits, as well as on the time resolution. However no visible permanent degradation of the performance due to the detector aging was observed. Efficiencies remained at the level of 98%, with spatial resolutions around 80 μm and time resolutions around 12 ns.

The hybrid Micromegas technology, using an additional GEM foil as preamplifier, has proved to be efficient to maintain Micromegas detector performance in conditions of large muon and hadron fluxes.

## 10. References

[1] Y. Giomataris, P. Rebourgeard, J.P. Robert and G. Charpak, MICROMEGAS: A high-granularity position-sensitive gaseous detector for high particle-flux environments, NIM A 376, 29 (1996)

[2] C. Bernet et al., The 40-cm x 40-cm gaseous microstrip detector Micromegas for the high-luminosity COMPASS experiment at CERN, NIM A 536 61-69 (2005)

[3] COMPASS Collaboration, COMPASS-II Proposal, SPSC-P-340, CERN-SPSC-2010-014, May 17, 2010

[4] COMPASS collaboration, COMPASS-II addendum 2, CERN-SPSC-2017-034, SPSC-P-340-ADD-1, April 5, 2018

[5] D. Neyret et al., New pixelized Micromegas detector for the COMPASS experiment, JINST 4 P12004 (2009)

[6] B. Adams et al., Letter of Intent: A New QCD facility at the M2 beam line of the CERN SPS (COMPASS++/AMBER), CERN-SPSC-2019-003 (SPSC-I-250), e-Print: 1808.00848 [hep-ex]

[7] F. Jeanneau et al., Performances and ageing study of resistive-anodes Micromegas detectors for HL-LHC environment, IEEE Nuclear Science Symposium and Medical Imaging Conference (NSS/MIC 2011), 23-29 Oct 2011. Valencia, Spain

[8] T. Alexopoulos et al., A spark-resistant bulk-micromegas chamber for high-rate applications, Nucl.Instrum.Meth. A640 (2011) 110-118

[9] F. Sauli, GEM: A new concept for electron amplification in gas detectors, Nucl.Instrum.Meth. A386 (1997) 531-534

[10] S. Kane et al., A study of MICROMEGAS with preamplification with a single GEM, proceedings to the ICATPP-7 conference October 15-19 2001, Villa Olmo, Como, Italy, DAPNIA-02-48

[11] S. Procureur et al., Origin and simulation of sparks in MPGD, JINST 7 (2012) C06009

[12] S. Procureur et al., Discharge studies in micromegas detectors in a 150-GeV/c pion beam, Nucl.Instrum.Meth. A659 (2011) 91-97

[13] G. Charles et al., Discharge studies in Micromegas detectors in low energy hadron beams, Nucl.Instrum.Meth. A648 (2011) 174-179

[14] M. Vandenbroucke, Development and Characterization of Micro-Pattern Gas Detectors for Intense Beams of Hadrons, PhD thesis defended July 2, 2012, TUM and Pierre&Marie Curie Universities

[15] S. Aune et al., Micromegas tracker project for CLAS12, Nucl.Instrum.Meth.A 604 (2009) 53-55, DOI: 10.1016/j.nima.2009.01.047

[16] F. Thibaud, Développement de détecteurs Micromegas pixellisés pour les hauts flux de particules et évaluation de la contribution diffractive à la leptoproduction de hadrons à COMPASS (In French), PhD thesis 29 September 2014, Université Paris-Sud, CERN-THESIS-2014-220

[17] M.J. French et al., Design and results from the APV25, a deep sub-micron CMOS front-end chip for the CMS tracker, Nucl. Instr. and Meth. A 466 (2001) 359

[18] F. Thibaud et al., Performance of large pixelised Micromegas detectors in the COMPASS environment, proceedings to MPGD 2013, 1-5 July 2013, Zaragoza, Spain, JINST 9 (2014) C02005

[19] COMPASS Collaboration (R. Akhunzyanov, et al.), Transverse extension of partons in the proton probed in the sea-quark range by measuring the DVCS cross section, Phys. Lett. B 793 (2019) 188-194, Phys.Lett.B 800 (2020) 135129 (erratum), DOI: 10.1016/j.physletb.2019.04.038

[20] C. Quintans for COMPASS coll., Future Drell-Yan measurements in COMPASS, J.Phys.Conf.Ser. 295 (2011) 012163, DOI: 10.1088/1742-6596/295/1/012163

[21] COMPASS Collaboration (M. Aghasyan, et al.), First measurement of transverse-spin-dependent azimuthal asymmetries in the Drell-Yan process, Phys. Rev. Lett. 119 (2017) 11, 112002, DOI: 10.1103/PhysRevLett.119.112002

[22] J.M. Le Goff et al., COMPASS plans to measure transversity, Nucl.Phys.A 711 (2002) 56-61, DOI: 10.1016/S0375-9474(02)01194-6

[23] COMPASS Collaboration (G. Alexeev, et al.), High-statistics measurement of Collins and Sivers asymmetries for transversely polarised deuterons, CERN-EP-2023-308 (2023), arXiv: 2401.00309 [hep-ex]

[24] Gianmaria Milani, Riccardo Longo, Angelo Maggiora, Daniele Panzieri, FLUKA simulation of the irradiation dose on the Compass PT material holder, COMPASS note 2013-10 (2013)

[25] A. Ferrari, P.R. Sala, A. Fasso, and J. Ranft, FLUKA: a multi-particle transport code, CERN-2005-10 (2005), INFN/TC_05/11, SLAC-R-773

## 11. Figures

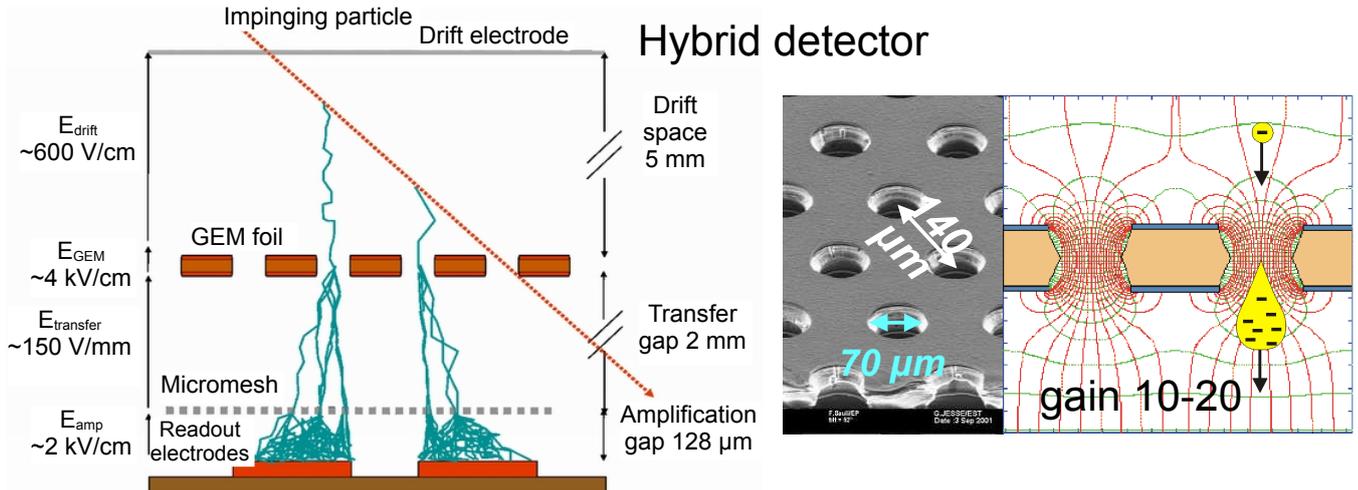

*Figure 1: Principle of hybrid Micromegas detectors. A GEM foil is placed 2mm above the micromesh to preamplify primary ionization electrons (left). A microscopic view of a GEM foil is shown with the electric field and potential lines in the foil holes (right)*

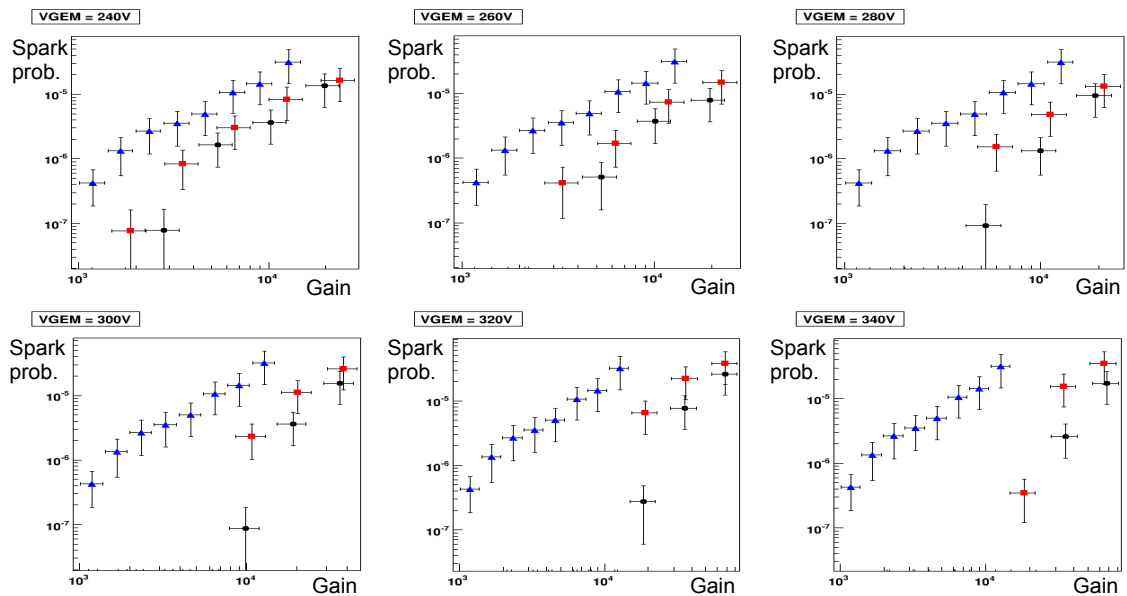

*Figure 2: Spark probability vs gain for a standard Micromegas (blue triangles), and for hybrid detectors at several GEM voltages. The GEM foil is placed at 1 mm (red squares) or 2 mm (black circles) above the micromesh.*

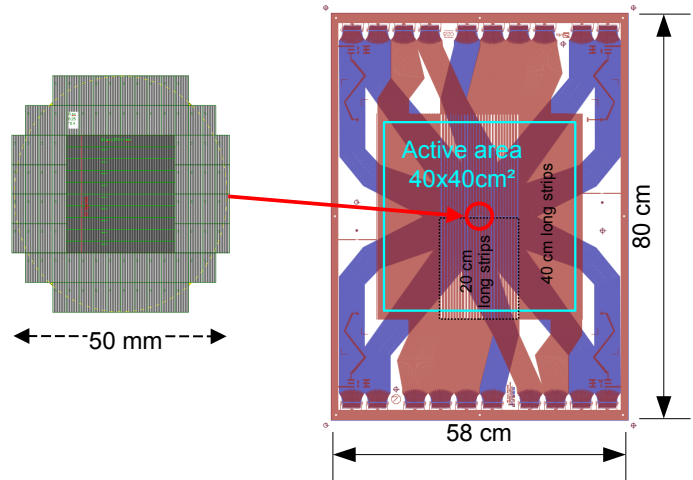

*Figure 3: Pixel geometry in the center of the detector, which covers a 5 cm-large area with 2.5x0.4 mm² pixels in the center and 6.25x0.4 mm² ones in the periphery (left). Geometry of the whole detector, with active area materialized by the cyan square (right). The red strips are on the top side of the readout board while the blue ones are on the bottom and connect the central pixels to the connectors on the edge of the detector.*

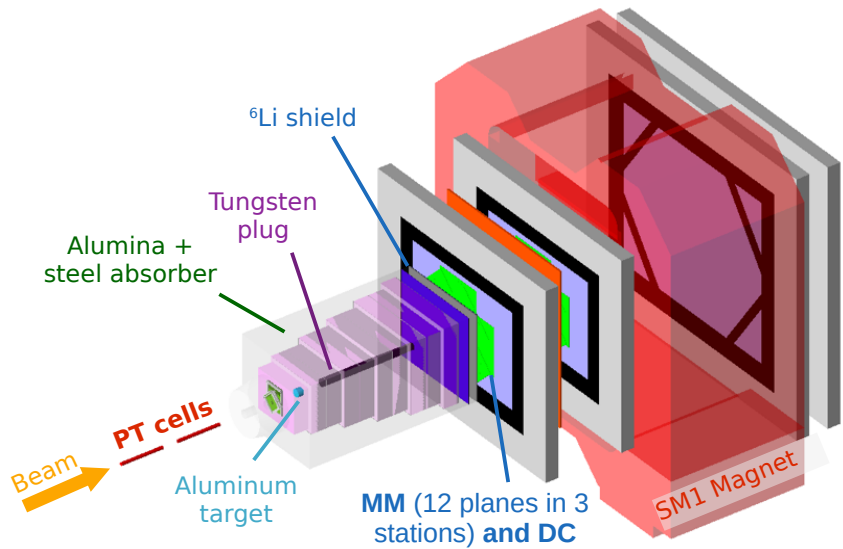

*Figure 4: COMPASS detector configuration in 2015 and 2018 during the Drell-Yan run. The Micromegas were placed right after the alumina and concrete absorber (the concrete part is not represented here). A tungsten plug was stopping the pion beam in the absorber. The lithium shield was not present in 2014-15*

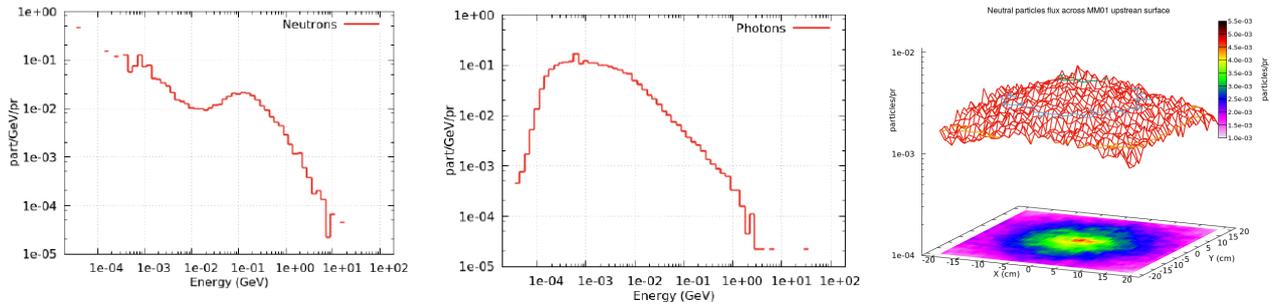

*Figure 5: Energy spectra of neutrons (left) and photons (center) emitted by the absorber and impinging on the first station of Micromegas detectors. Neutral particle flux over the first Micromegas detector surface (right). Plots from Angelo Maggiora (unpublished results)*

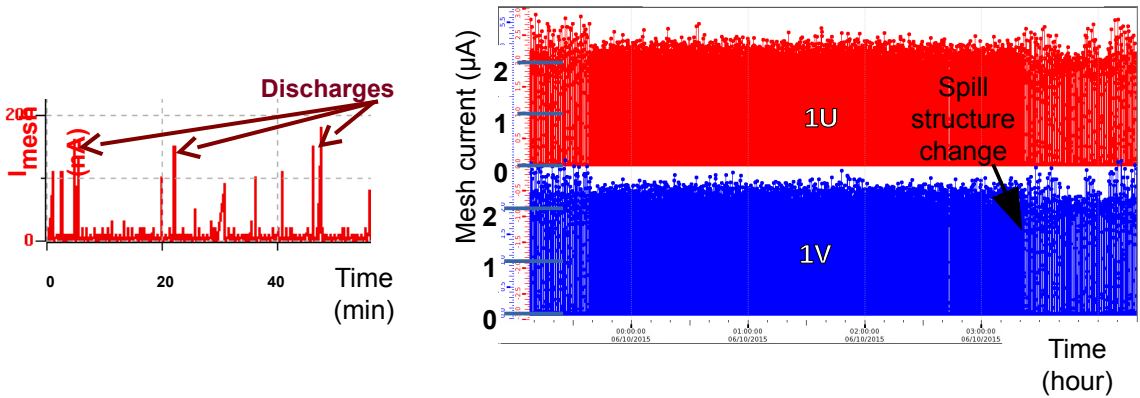

*Figure 6: Mesh current of a former Micromegas detector (left), the discharges are clearly visible every 5-10 minutes. Mesh current of two hybrid detectors of station 1 (right), with amplification currents larger than 2 µA, the discharges are barely visible*

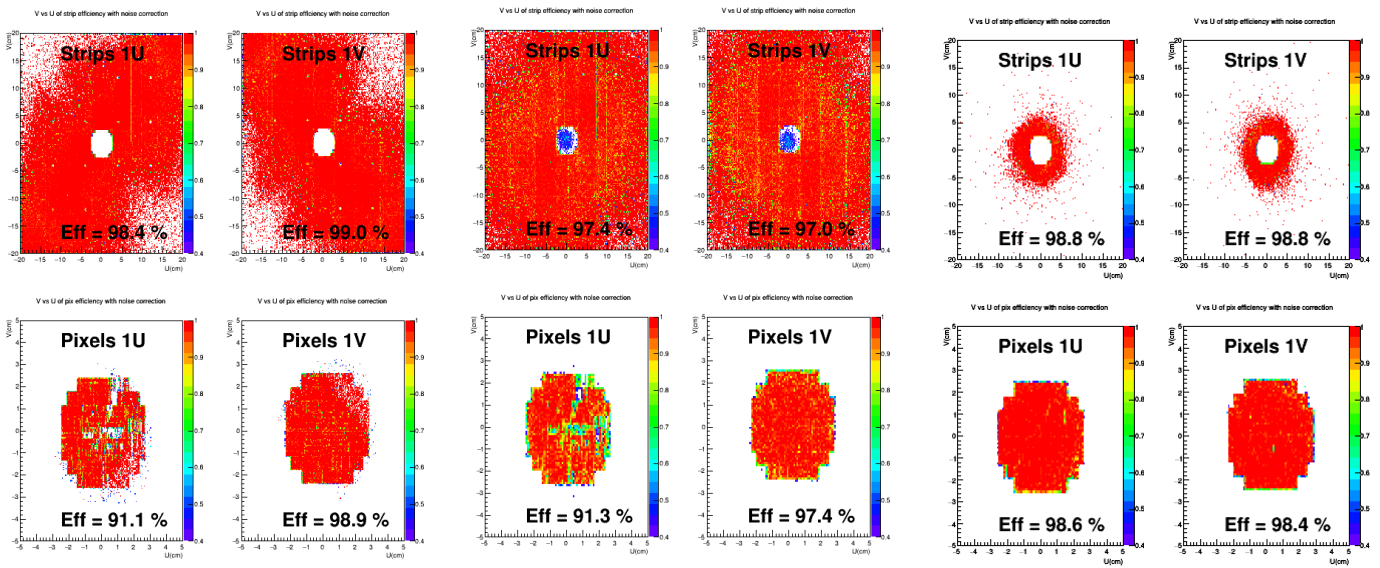

*Figure 7: Pseudo-efficiency of U and V detectors of station 1, for strip (top) and pixel (bottom) areas. The 4 left plots show the efficiencies with low intensity muons during the 2015 Drell-Yan run, giving the intrinsic detector efficiencies. The 4 center plots are taken with large intensity pion beam. The 4 right plots were taken during the 2016 DVCS run with medium intensity muon beam.*

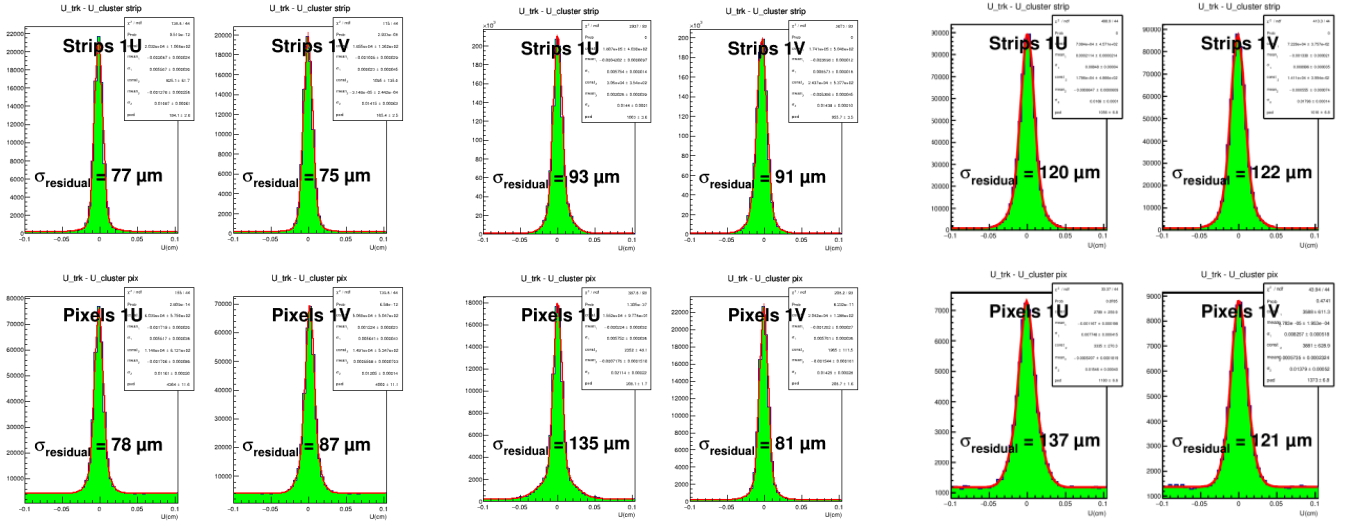

*Figure 8: Spatial residuals of U and V detectors of station 1, for strip (top) and pixel (bottom) areas. The 4 left plots show the residuals during the 2016 DVCS run with medium intensity muon beam. The 4 center plots are taken during the Drell-Yan run with low intensity muon beam, while the 4 right plots are taken during the same run with high intensity pion beam.*

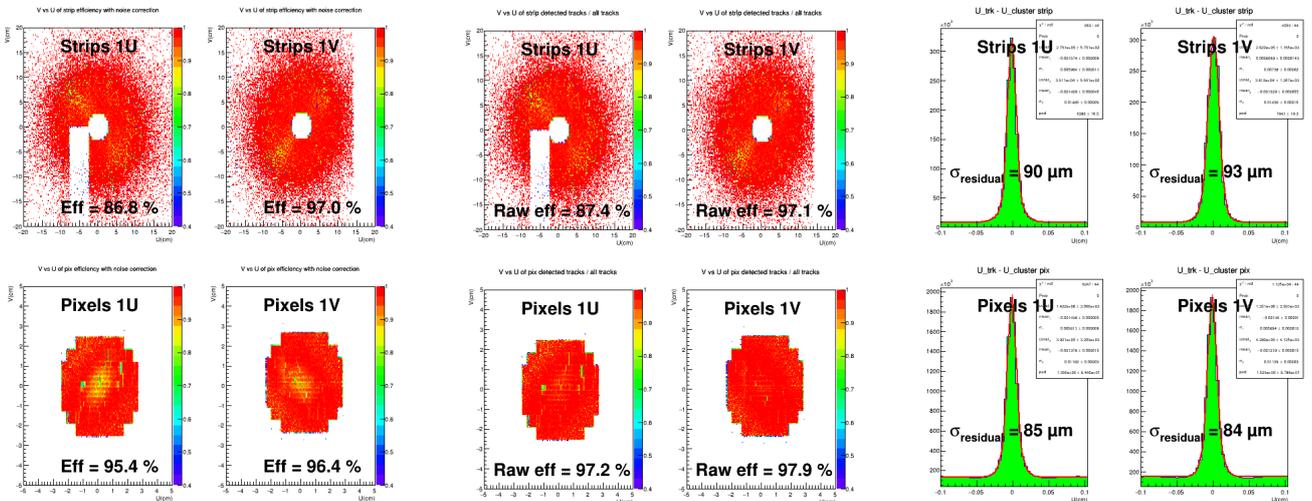

*Figure 9: Pseudo-efficiency of U and V detectors of station 1 (4 left plots) during the 2022 transversity run with high intensity muon beam. Two APV readout cards were not functioning during that period. Same efficiency in same conditions without the correction of pile-up effects (4 center plots). Spatial residuals of the same detectors in the same conditions (4 right plots).*